\documentclass{emulateapj}

\begin{document}

\title{A Catalog of 24\,$\micron$ Sources Toward the Galactic Center}

\author{J.~L. Hinz\altaffilmark{1}, G.~H. Rieke\altaffilmark{1}, F. Yusef-Zadeh\altaffilmark{2}, J. Hewitt\altaffilmark{2}, Z. Balog\altaffilmark{1,3}, M. Block\altaffilmark{1}}

\altaffiltext{1}{Steward Observatory, University of Arizona, 933 N.\ Cherry Ave.,  Tucson,
AZ  85721}

\altaffiltext{2}{Department of Physics and Astronomy, Northwestern University, 2145 Sheridan Rd., Evanston, IL  60208}

\altaffiltext{3}{On leave from the Department of Optics and Quantum Electronics, University of Szeged, H-6720, Szeged, Hungary}

\begin{abstract}

We present a $\sim1.5\arcdeg\times8\arcdeg$ (220$\times$1195\,pc)
Multiband Imaging Photometer for 
{\it Spitzer} 24\,$\micron$ image of the Galactic Center and an
accompanying point source list.  This image is the highest spatial resolution
(6$\arcsec \sim 0.25$\,pc) and sensitivity map ever taken across the 
GC at this wavelength, showing
the emission by warm dust in unprecedented detail.  Over 120,000 point sources
are identified in this catalog with signal-to-noise ratios greater than
five and flux densities from 0.6\,mJy to 9\,Jy.

\end{abstract}

\keywords{Galaxy: center - infrared: stars}

\section{INTRODUCTION}

The Galactic Center (GC) has long been a source of great interest at 
infrared (IR) wavelengths due to the complex, obscured nature of the region
and its areas of energetic, massive star formation (Becklin \& Neugebauer 
1968; Rieke et al.\ 1978).  Early experiments at far-IR wavelengths
produced maps with resolution on the order of arcminutes, offering some of the 
first looks into the inner workings of the Galaxy (e.g., Odenwald \& Fazio 
1984; Campbell et al.\ 1985).

As IR technology improved dramatically, a new generation of images became 
available, and two large surveys were launched.  The first was 
a map of the inner galaxy conducted with {\it Infrared 
Space Observatory} ({\it ISO}; Kessler et al.\ 1996) at 7 and 15\,$\micron$.
These data had a resolution of 6$\arcsec$ or better and were combined
with DENIS $IJK_{\rm s}$ data to quantify the spatial distributions of 
the stellar populations and their properties in the GC and to determine the 
interstellar extinction in that region.  This ISOGAL project 
(Omont et al.\ 2003) had
nearly complete point-source detection down to  $\sim$\,10-20\,mJy at 
7\,$\micron$ and $\sim$10\,mJy at 15\,$\micron$, subject to issues of
crowding and background levels in some areas.  The observed regions 
($\sim$16 deg$^2$) were distributed 
along the inner Galactic Disk, mostly within 
$\vert \ell \vert<30^\circ$, $\vert {\it b} \vert<1^\circ$.
The second survey was that of the Galactic plane within $|b|\le5\arcdeg$ 
(Price et al.\ 2001), carried out with the {\it Midcourse Space Experiment}
({\it MSX}; Mill et al.\ 1994) at a spatial resolution of 
$18\farcs3$\,pixel$^{-1}$ and with bands sensitive from $\sim$6-25\,$\micron$.

Extending these works with the new capabilities of the Multiband
Imaging Photometer for {\it Spitzer} (MIPS; Rieke et al.\ 2004), we present a 
$\sim1\arcdeg5\times8\arcdeg$ (220$\times$1195\,pc, assuming a distance of
8.5\,kpc) scanmap of the GC at
24\,$\micron$ and an associated point source catalog for $|l| < 4\arcdeg$ and 
$|b| < 1\arcdeg$.  The efficiency of
the mapping and the high sensitivity of the instrument allow a quick and deep
map of the GC to be produced that outperforms previous missions.
For comparison, the {\it MSX} E passband at 21.34\,$\micron$, with 
an isophotal bandwidth of 6.24\,$\micron$, is similar to the
MIPS 24\,$\micron$ window.  The {\it MSX} noise 
equivalent radiance in this wavelength range is approximately 
40\,MJy\,sr$^{-1}$ for the inner galaxy and 65\,MJy\,sr$^{-1}$ for the 
outer galaxy, with a sensitivity of $\sim\,3000$mJy.  The MIPS sensitivity is 
0.22\,MJy\,sr$^{-1}$, 
corresponding to a point source limit for low background of 212\,$\mu$Jy.  
With complex backgrounds, the order of magnitude smaller beam area with MIPS 
results in larger gains.
However, due to its lower sensitivity, almost no sources along the galactic 
plane are saturated in the {\it MSX} data set, making it complementary
to the new, deeper observations.  For a comparison of the sensitivity and
saturation levels of 2MASS, IRAC, {\it MSX}, and MIPS 24\,$\micron$, see Table
1 of Robitaille et al.\ (2007), and see their Figure 1 for spectral response 
curves of filters for those instruments.

Our data are also complementary to the {\it Spitzer} Legacy program
MIPS Inner Galactic Plane Survey 
(MIPSGAL; P.I.\ S.\ Carey) which observed 220 deg$^2$ of the inner 
Galactic plane, 
$65\degr > l > 10\degr$ and $-10\degr > l > -65\degr$ for $|b| < 1\degr$, at 24 and 
70\,$\micron$ with MIPS
(Carey et al.\ 2005, 2006; Shenoy et al.\ 2007) but avoided the GC.  There
is some mild scanning overlap between our observations and those of MIPSGAL
at the edges of each survey, but together they provide a large, comprehensive
overview of this region.

A smaller survey of the GC has been conducted with the {\it Spitzer's}
Infrared Array Camera (IRAC; Fazio et al.\ 2004), covering the 
inner $2\arcdeg\times1.5\arcdeg$ ($\sim300\times220$\,pc)
and much of the Central Molecular Zone of the Galaxy
with the four bandpasses at 3.6, 4.5, 5.8, and 8.0\,$\micron$ (Program ID 3677,
P.I.\ Stolovy; Stolovy et al.\ 2006; Ramirez et al.\ 2008; Arendt et al.\ 2008).  Future
work will combine their point source catalog with ours, and other more 
detailed results about the nature of the sources are presented elsewhere
(Yusef-Zadeh et al.\ 2008; Yusef-Zadeh et al.\ 2008, in preparation).  Table 1
summarizes the relevant parameters of the various GC region surveys.


\section{OBSERVATIONS}

The MIPS observations of the GC (Program ID 20414) were obtained in fast scan 
mode with $2\arcdeg$ scan legs and with full array ($302\arcsec$) offsets.  
Eight astronomical observation requests (AORs), each containing twelve of 
these scan legs, were designed 
symmetrically about the GC and were obtained at an angle of $58.6\arcdeg$ from 
the plane for a field size of $\sim 1\arcdeg5\times8\arcdeg$.  The entire map, 
at a scale of $2\farcs55$, is shown in Figure 1.  The fast scan observing mode 
was chosen due to 
the brightness of the GC itself, to avoid saturating completely over
the plane, resulting in a total MIPS mapping time of $\sim$\,18 hours.  
Despite the short exposure, the inner $\sim$\,100 square arcminutes 
are saturated, as are 
many individual sources throughout the image, especially along
the plane.

The operation of MIPS in scan-map mode allows for simultaneous
mapping at all three wavelengths.  However, the use of full array
offsets means that the 70\,$\micron$ 
data from the survey, with only half of the array working,
are incomplete and, moreover, are laden with scan-related artifacts due
to crossing the galactic plane multiple times, so the scan map has
limited value, especially near the central region.  The 160\,$\micron$ map is 
completely saturated.

Starting with raw data from the {\it Spitzer} Science Center (SSC),
data reduction was performed using version 3.02 of the MIPS Data 
Analysis Tool (DAT; Gordon et al.\ 2005).  The DAT performs standard
processing of infrared detector array data (e.g., dark subtraction, flat
fielding) as well as steps specific to the MIPS arrays (droop correction).
The data were then additionally processed with custom routines
specifically written for scan map data sets with bright sources,
such as the SAGE program to observe the Large Magellanic Cloud (LMC;
Meixner et al.\ 2006) and the GTO program to observe M\,31 (Gordon
et al.\ 2006).  In these routines, possible readout offset was corrected
because one of the four MIPS readouts drifts slightly.  The processing 
excludes images affected by saturating 
sources, e.g., persistence that appears in a frame after a source has 
saturated the detector, and corrects for artifacts associated with the
direction of scan.  The first four data collection events in each scan 
leg are excluded because of transients associated with the boost frame. 
Two of the eight scan legs (AORs 14657280 and 14657536) were observed six 
months apart from the others.  The difference in the background level due to 
changes in the zodiacal light between the two sets of observations was 
subtracted within the custom programming.  Due to the complicated nature of
diffuse emission in the Galactic Center region, all other background 
subtraction was performed on individual sources locally as photometry was
extracted (see section 3).
The entire mosaic was calibrated using the most recent values available for
24\,$\micron$ data (Engelbracht et al.\ 2007).

\section{CREATION OF THE SOURCE CATALOG}

To create a source catalog, we used the StarFinder program, an IDL 
code created for deep analysis of stellar fields (Diolaiti et al.\ 2000).  
The image mosaic was broken into approximately 50 tiles, most of which
are $\sim1000\times1000$ pixels in size, to speed efficiency of the StarFinder
software.  StarFinder can be used such that it extracts the point spread
function (PSF) directly from the images to take into account the actual 
structure of the instrumental response.  However,
in this case, the PSF is not taken from the image itself, but generated
using the STinyTim (Krist 2002) model, as in the case of the SAGE project.
The PSF of the MIPS instrument is very stable and using the STinyTim model
allows us to avoid re-deriving a PSF for every new data set (see Gordon
et al.\ 2007 for discussion of MIPS PSF comparisons).
Objects were constrained to have a FWHM of $\sim5\arcsec$, a
signal-to-noise threshold of 5, and a minimum correlation of 0.9.
The correlation parameter indicates an object's match to the input
PSF.  Higher numbers indicate closer matches, with 1 being perfect.
Photometry was performed by StarFinder, where a background box size of
ten times the FWHM of the PSF is specified.
Table 2 contains a truncated source list for the GC, including (1) 2MASS identification
of the source, if any (see Section 4.1), (2) RA, (3) DEC, (4) calculated error 
in RA, (5) calculated error in DEC, (6) flux density at 24\,$\micron$, (7) 
error in the flux density, and (8) the correlation.  These errors are the
Starfinder nominal fitting errors only.  More representative uncertainties are
of order 7\% for the flux densities and $0\farcs6$ for the positions (see the
discussion in Section 4.1).
The entire source list can be accessed via the 
electronic edition of this article.

There are some obvious difficulties in producing a usable point source
catalog for this data set.  The first is that there are
many saturated areas in the image.  For instance, the inner 
$\sim10\arcmin\times10\arcmin$,
including the well known Arches and Quintuplet Cluster regions, are 
saturated, along with several regions of bright diffuse emission along the 
galactic plane.  Point sources detected along the edges of these regions 
may be false or may have their flux densities
influenced by areas of unusually bright emission in the diffuse background
(background estimates generated by the SSC's SPOT run from $\sim$100-300 MJy 
sr$^{-1}$ in the GC region).
Especially near the largest continuous saturated area in the center of
the mosaic, all flux densities for point sources in the catalog within 
2$\arcmin$ of the saturated edge should be treated with caution.
An example of this effect is shown in Figure 2,
where a region of diffuse emission near the Galactic Center is shown.  The 
MIPS image on the left shows real sources embedded in bright diffuse emission, 
while the sources identified by StarFinder on the right show false bright 
point sources, e.g., at the very center of the image and all along the bright
edge of the diffuse emission.

Individual objects scattered about the mosaic that are saturated at 
24\,$\micron$ are not found by StarFinder and so are not included in 
Table 2.  StarFinder occasionally locates points in the Airy pattern around
very bright or saturated point sources and adds them to the
catalog as if they were real, individual sources.  Figure 3 shows a typical 
example of this
phenomenon around a saturated source at 17h53m18.8s, -26d56m37.2s J2000.  On
the left is the reduced MIPS image, and on the right is an image created
by StarFinder of all the objects it identifies in the field.  The
central, saturated object is not identified by StarFinder as a source, but the 
ring associated with the MIPS instrument pattern is incorrectly identified as 
several individual point sources.  Due to
the large area covered by the map, it is not feasible to remove by hand each of
these false sources from the catalog (although many were removed in
this manner for tiles along the plane).  For this reason, all sources
within a radius of 30\,$\arcsec$ of objects with fluxes above $2$\,Jy, just 
outside the Airy ring, are possibly false.  Though some of the sources detected
within this radius may be real, we caution that even real sources close to
bright or saturated sources likely have unreliable flux densities affected
by the bright neighbor. 

\section{DISCUSSION}

\subsection{Calibration and Positional Accuracy}
Objects in the
source list have an average flux density uncertainty given by StarFinder 
of $\sim$3\%.
In addition to this, photometric calibration uncertainties for 
MIPS 24\,$\micron$ data are estimated to be 4\% (Engelbracht et al.\ 2007),
dominated by uncertainties in the absolute calibration. 
Figure 4 shows the flux density distribution of sources in the 
catalog.

The average positional uncertainty for objects cataloged by 
StarFinder is $0\farcs1$.  This value represents only nominal fitting errors,
and a more representative positional uncertainty for MIPS 24\,$\micron$ data
is $\sim\,0\farcs6$ (e.g., Meixner et al.\ 2006).  
Positions for sources were compared and cross-correlated with the Two Micron 
All Sky Survey (2MASS; Skrutskie et al.\ 2006) to test for possible distortion
effects or other systematic problems.  Over the entire field, we find an 
average systematic offset of +1$\farcs$6 in right ascension and +0$\farcs$5 in 
declination between the 2MASS and 24\,$\micron$ sources.  This offset in right 
ascension is larger than normally found between 24\,$\micron$ and 2MASS data, 
though the reason is unclear.  There is also some variation in the offset from 
tile to tile, particularly in right ascension.  Figure 5 shows this variation 
across the field in uncorrected coordinates. 

Given the variations, we therefore performed coordinate corrections for each 
of the 50 image tiles individually.  Each object in the 24\,$\micron$ catalog
was matched with the closest neighbor in the 2MASS catalog within 0.0012\,deg
($4\farcs32$).  A coordinate shift between the 2MASS and 24\,$\micron$ data
was determined using a constant in right ascension and in declination for each 
of the 50 tiles.  The 24\,$\micron$
source coordinates were adjusted according to the individual calculated shifts;
these adjusted coordinates appear in Table 2.  The rms positional uncertainty
after this adjustment is $0\farcs58$ in RA and $0\farcs61$ in DEC.  Matching 
between the two
catalogs was conducted again using a reduced radius of 0.0006\,deg 
($2\farcs16$; similar resolution to the 2MASS map).  The nearest neighbor 
within the matching radius was identified
as well as all neighbors within the matching radius.  Less than 2\% of the
24\,$\micron$ sources have multiple 2MASS matches in this radius.  The 2MASS
matches are listed in Table 2.

\subsection{Intruders and Chance Identifications}

The main purpose in assembling the source catalog is to expedite Galactic 
studies. We define intruders as sources that are within the Solar System or 
outside the Milky Way. Two additional classes are spurious detections, for 
example due to cosmic ray hits on the detectors, and ghosts and other 
consequences of nearby strong sources or strongly emitting extended regions. 
The issue of spurious sources is largely managed by the high level of 
redundancy in the MIPS 24\,$\micron$ data-taking. For this program, each 
source was observed ten times as it crossed the detector array on one 
scan map leg, and then ten more times as it crossed in the opposite direction 
on the next leg. Ghosts and similar issues have been discussed in Section 3. 

One form of intruder is extragalactic sources. From, e.g., Papovich et 
al.\ (2004), the average counts for 24\,$\micron$ sources brighter than 1\,mJy 
is about 600 per square degree, for sources brighter than 5\,mJy it is about 
30 per square degree, while for sources brighter than 10\,mJy it is 
about 25 per square degree. The total source count at 24\,$\micron$ in the 
Galactic Center is about 10000 per square degree, most brighter than 5\,mJy. 
Thus, the extragalactic 
contamination is likely to be at the $\sim$ 0.3\% level. Although this low level
sounds benign, programs isolating sources with extreme properties need to be 
pursued with caution to avoid extragalactic objects (most of the extragalactic 
sources will be below the 2MASS detection limits, for example). 

The Galactic Center is at low ecliptic latitude ($\sim6\arcdeg$), so there 
will be a significant population of asteroids projected onto it. The 
24\,$\micron$ map was made in a single visit, and the time between scan legs 
was inadequate to identify asteroids by their motion. From typical asteroid 
counts as a function of ecliptic latitude (Hines et al.\ 2007; D.~C. Hines, 
private communication, 2008), we expect up to $\sim$ 1000 per square degree 
brighter 
than 1\,mJy and $\sim$ 20 per square degree brighter than 10\,mJy. Again, compared 
with the total source counts, the implied level of contamination is 
$\le$ 0.3\%, but 24\,$\micron$-only detections should be viewed with caution 
until they have been confirmed through detection in some other way, 
e.g., by IRAC at 5.8 and 8\,$\micron$ (Ramirez et al.\ 2008). 

Another form of false result is association of a 24\,$\micron$ source with a 
random object not physically associated. An example is our identification 
of sources in our catalog with those in 2MASS. For 2MASS, we take the 
positional errors to be $\sim0\farcs1$ (Cutri et al.\ 2003; 
Zacharias et al.\ 2003).  The 
errors are therefore dominated by those at 24\,$\micron$, and we have allowed 
positional discrepancies up to $1\farcs5$. In two 40 $\times$ 40 arcmin 
reference fields, we found 60,000 2MASS sources and 4,000 24$\mu$m ones. 
If the two source catalogs were uncorrelated, the probability of a randomly 
associated 2MASS object for any given 24\,$\micron$ one would be $\sim$ 15\%. 
However, the catalogs are strongly correlated and a better estimate of the 
number of false identifications is $\sim 2\%$, based on the cases with two 
2MASS sources within the matching radius.  Our adopted tolerance is about 
2.5 times the expected typical ($\sim 1 \sigma$) 24\,$\micron$ positional 
errors, so the number of false associations with 2MASS sources can be 
reduced significantly by requiring slightly more accurate positional 
agreement, at the cost of rejecting a modest number of correct associations. 

The IRAC catalog of a $2\arcdeg\times1.4\arcdeg$ region (Ramirez 
et al.\ 2008) is another logical set of sources for band merging, 
particularly at 5.8 and 8\,$\micron$ where many of the 24\,$\micron$ sources 
should be detected independently. At 8\,$\micron$, the scale 
is $1\farcs2$, and there are roughly 110,000 sources/square degree (Ramirez et 
al.\ 2008). The probability of a random association of an 8\,$\micron$ source 
with one at 24\,$\micron$ is then $\sim$ 25\%. The other IRAC bands need to 
be used as part of band merging, since they both have more accurate 
positions and can provide spectral information to confirm the plausibility 
of any associations.  These complications arise because the IRAC catalog 
is confusion limited. As a result, we have not attempted to cross identify 
with the IRAC sources. The optimum way to do so will depend on the 
specifics of a program and the resulting expectations for the mid-infrared 
spectral energy distributions of its target objects. 

\subsection{What is in the Catalog?}

On the Vega system, 11.3\,mJy at 24\,$\micron$ corresponds to 7th magnitude. 
Detections in 2MASS and with IRAC are typically confusion limited at 
12 - 13th magnitude. At the Galactic nucleus itself, where A$_V$ $\sim$ 30, 
these detection limits correspond to reddening corrected colors of 
K$_S$ - [24] $\sim$ 3. The extinction over the mapped region is highly 
variable; it is much greater than 30 magnitudes just to the SE of the 
nucleus because of the projection of the Galactic plane onto this region, 
and it is substantially less over much of our map out of the plane. 
Nonetheless, it is noteworthy that over significant regions, the 
24\,$\micron$ data are deep enough to identify sources with modest 
K$_S$ - [24] excesses. The distance modulus to the Galactic Center 
is m-M = 14.4. Therefore, m$_{K_S}$ = 13 corresponds to M$_{K_S}$ = -1.4 
for no extinction and to M$_{K_S}$ = -4.8 for A$_V = 30$ (assuming the 
extinction law of Rieke \& Lebofsky 1985). Normal giant stars of type 
G8 and later have M$_{K_S}$ $\le$ -1.4, while giants of type M5 and 
later and supergiants of virtually all types have M$_{K_S}$ $\le$ -4.8 
(Lang 1992;
Tokunaga 1999). Therefore, the dominant objects in the catalog are red 
giants and supergiants with 24\,$\micron$ excess emission. Extreme forms of 
mass losing late-type stars will also be well represented, e.g., OH/IR 
stars (with many examples in the Galactic Center region being at 
$\sim$\,1\,Jy at 20\,$\micron$ and with K - Q from 5 to 10, Blommaert 
et al.\ 1998). 

There will be additional classes of source, many below the 2MASS detection 
limits. We can estimate roughly how many such objects to expect. Of the 
60,000 objects brighter than 10\,mJy roughly 25\% or 15,000 do not 
have 2MASS identifications. Taking the false identification rate to be 
2\%, there are another 900 such sources for a total of 15,900. Approximately 
540 of these sources are likely to be extragalactic or asteroids; the 
remaining must represent extreme mass losing and embedded 
post-main-sequence stars, ultra-compact H\,{\sc ii} regions and other 
products of 
recent star formation, planetary nebulae, and supernova remnants. The 
latter two categories will be rare. Only a few hundred planetary nebulae 
are expected in this region (Jacoby \& Van de Steene 2008). Supernova 
remnants are also relatively uncommon and may not always be strong infrared 
emitters (e.g., Reach et al.\ 2006).  A large number of candidate young
stellar objects have been identified and will be explored in a future paper
(Yusef-Zadeh et al.\ 2008, in preparation).

\subsection{Source Distribution and Extended Emission}

The 24\,$\micron$ sources are concentrated along the
Galactic plane with a scale height similar to the distribution of molecular
gas in the GC region ($\sim0.5\degr$; Bally et al.\ 1988). There is an excess 
of $\sim30000$
sources at negative longitudes.  This asymmetry could be explained by the
high extinction experienced by mid-IR sources in the region where dense
molecular clouds are highly concentrated.  Alternatively, the distribution of
compact dusty sources could be intrinsic due to an excess of young stellar
objects or evolved asymptotic giant branch stars.  These possibilities will
be discussed in Yusef-Zadeh et al.\ (2008, in preparation).

The large scale view of the surveyed region at 24\,$\micron$ shows that the 
central
sources distributed between -1.8$\degr< l < 0.8\degr$ and $|b|<0.9\degr$
are the brightest. The mean brightness of this region is roughly 4 to 5 times
higher than the region beyond the inner few degrees of the Galactic
center. Several extended H\,{\sc ii} regions detected at 24\,$\micron$ 
are identified on the basis of radio
continuum and submillimeter counterparts (Yusef-Zadeh et al.\ 2004; 
Pierce-Price et al.\ 2000).
H\,{\sc ii} complexes along the Galactic plane at 24\,$\micron$ are
associated with Sgr A -- E and the radio arc. At positive longitudes, a
string of infrared dark clouds (IRDCs) is concentrated between
l$\sim0.2\degr$ and Sgr B2 near l=0.7$\degr$. Submillimeter emission from
these clouds is prominent (Lis \& Carlstrom 1994; Pierce-Price et al.\ 2000). At negative
latitudes, the distribution of dust emission at 24\,$\micron$ shows that the
IRDCs associated with the 20 and and 50\,km\,s$^{-1}$ molecular clouds 
M-0.13-0.08
and M-0.02-0.07, are detected, respectively (Herrnstein \& Ho 2005; Armstrong
\& Barrett 1985). Both of these clouds are known to
be located near the GC. At positive latitudes, two extended
clouds known as the western and eastern GC lobes are shown
prominently near l $\sim -0.5\degr$ and $\sim 0.2\degr$, respectively. At
negative latitudes, there are several counterparts to foreground
H$\alpha$ emission line nebulae which are prominent at 24\,$\micron$.

\section{SUMMARY}

A new MIPS mosaic of the GC at 24\,$\micron$ reveals the warm dust
emission in the area at unmatched sensitivity and spatial resolution.
A point source catalog for objects in the inner $\sim1\arcdeg5\times8\arcdeg$
has been produced, containing over 120,000 candidate sources between 0.6\,mJy 
and 9\,Jy.
Of these sources, $\sim540$ are likely to be intruder sources not associated with the 
GC, such as extragalactic objects or asteroids.  These data are complementary 
to other surveys of this region and should
fuel multi-wavelength comparisons and studies of the inner Galaxy.

\acknowledgments

This work is based on observations made with the {\it Spitzer Space
Telescope}, which is operated by the Jet Propulsion Laboratory, California
Institute of Technology, under NASA contract 1407.
This publication makes use of data products from the Two Micron All Sky 
Survey, which is a joint project of the University of Massachusetts and the 
Infrared Processing and Analysis Center/California Institute of Technology, 
funded by the National Aeronautics and Space Administration and the National 
Science Foundation.

\clearpage

\begin{deluxetable}{llll}
\tablecaption{Surveys of the Galactic Center Region}
\tablewidth{300pt}
\tablehead{
\colhead{Survey} & \colhead{Coverage} & \colhead{Wavelength} & \colhead{Resolution} \\
\linebreak & (deg$^2$) & ($\micron$) & (arcsec)}
\startdata
2MASS &	All Sky	&	1.25	& 2 \\
2MASS &	All Sky	&	1.65	& 2 \\
2MASS &	All Sky	& 	2.15	& 2 \\
IRAC1 &	3	&	3.6	& 2 \\
IRAC2 &	3	&	4.5	& 2 \\
IRAC3 &	3	&	5.8	& 2 \\
IRAC4 &	3	&	8.0	& 2 \\
ISO1  &  16     &         7     & 6 \\
ISO2  &  16     &         15    & 6 \\
MSXA  &	3600	&	8.28    & 18 \\	
MSXC  &	3600	&	12.13	& 18 \\
MSXD  &	3600	&	14.65	& 18 \\
MSXE  &	3600	&	21.3	& 18 \\
MIPS  &	12	&	23.8	& 6 \\
\enddata
\end{deluxetable}

\begin{deluxetable}{llllllll}
\tabletypesize{\scriptsize}
\tablecaption{Point Source Catalog at 24\,$\micron$}
\tablewidth{0pc}
\tablehead{
\colhead{2MASS ID} & \colhead{RA} & \colhead{DEC} & \colhead{$\sigma_x$} & \colhead{$\sigma_y$} & \colhead{$F_{24}$} & \colhead{$\sigma_{F_{24}}$} & \colhead{Corr.} \\
\linebreak & (deg) & (deg) & (arcsec) & (arcsec) & (Jy) & (Jy) \\
\linebreak (1) & (2) & (3) & (4) & (5) & (6) & (7) & (8) }
\startdata\nodata          &  266.287841 & -30.977419 &  0.02927 &  0.0291 &  3.551 &
  0.02323 &  0.9679 \\
 17434488-3044340 &  265.937084 & -30.742722 &  0.02918 &  0.02632 &  3.532 &
  0.02271 &  0.9486 \\
 17435853-3049345 &  265.993959 & -30.826151 &  0.02939 &  0.02374 &  3.187 &
  0.02148 &  0.9867 \\
 \nodata          &  266.14624 & -30.438144 &  0.02468 &  0.01924 &  3.169 &
  0.02009 &  0.9799 \\
 17434410-3056446 &  265.933789 & -30.945665 &  0.01518 &  0.02561 &  2.312 &
  0.01703 &  0.9852 \\
 \nodata          &  266.255206 & -31.209862 &  0.02478 &  0.02745 &  1.998 &
  0.01255 &  0.9523 \\
 \nodata          &  265.761335 & -30.949127 &  0.02335 &  0.02361 &  1.973 &
  0.009045 &  0.9953 \\
 17443502-3034437 &  266.14593 & -30.578774 &  0.0172 &  0.01993 &  1.953 &
  0.01406 &  0.9759 \\
 17441335-3056200 &  266.056184 & -30.939233 &  0.01247 &  0.006901 &  1.776 &
  0.008141 &  0.9562 \\
 17451342-3038180 &  266.305943 & -30.638432 &  0.02665 &  0.02199 &  1.309 &
  0.009429 &  0.9831 \\
 17441640-3039103 &  266.068336 & -30.65275 &  0.0258 &  0.02424 &  1.256 &
  0.007597 &  0.9847 \\
 17434905-3041136 &  265.954328 & -30.687175 &  0.02936 &  0.02643 &  1.139 &
  0.006561 &  0.99 \\
 17435511-3045401 &  265.979572 & -30.760982 &  0.02681 &  0.02864 &  1.1 &
  0.007661 &  0.9764 \\
 17432702-3051042 &  265.862481 & -30.851173 &  0.0305 &  0.02928 &  1.098 &
  0.007793 &  0.9778 \\
 17452158-3052455 &  266.340014 & -30.879241 &  0.02936 &  0.02879 &  1.078 &
  0.007192 &  0.9874 \\
 17443132-3048224 &  266.130579 & -30.806163 &  0.01412 &  0.02999 &  1.039 &
  0.007664 &  0.9821 \\
 17450523-3054496 &  266.27189 & -30.913826 &  0.02931 &  0.02925 &  0.9833 &
  0.006514 &  0.9806 \\
 17450768-3035121 &  266.281693 & -30.587087 &  0.02922 &  0.03381 &  0.8739 &
  0.0068 &  0.9738 \\
 17441761-3056218 &  266.073363 & -30.939407 &  0.02717 &  0.02896 &  0.8129 &
  0.005895 &  0.9722 \\
 17434355-3050521 &  265.931548 & -30.847771 &  0.02747 &  0.03038 &  0.7588 &
  0.005271 &  0.9776 \\
\enddata
\tablecomments{For the
full list of over 120,000 sources, see the online version of this paper.}
\end{deluxetable}

\clearpage

\begin{figure}
\plotone{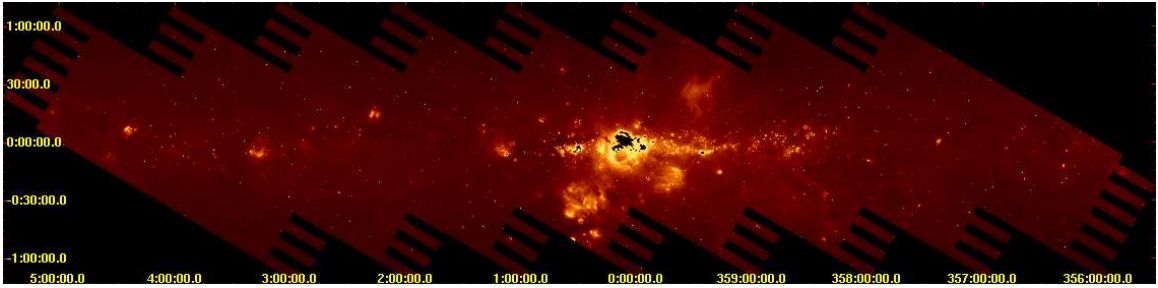}
\caption{The full MIPS map of the GC at 24\,$\micron$ at a scale of
$2\farcs55$.  Data are shown with the galactic plane along the horizontal,
an angle of $-58.7\arcdeg$ from north up and east to the left.  The approximate
coverage is $1.5\arcdeg\times8\arcdeg$.  The image is shown with a log stretch.}
\end{figure}

\begin{figure}
\plotone{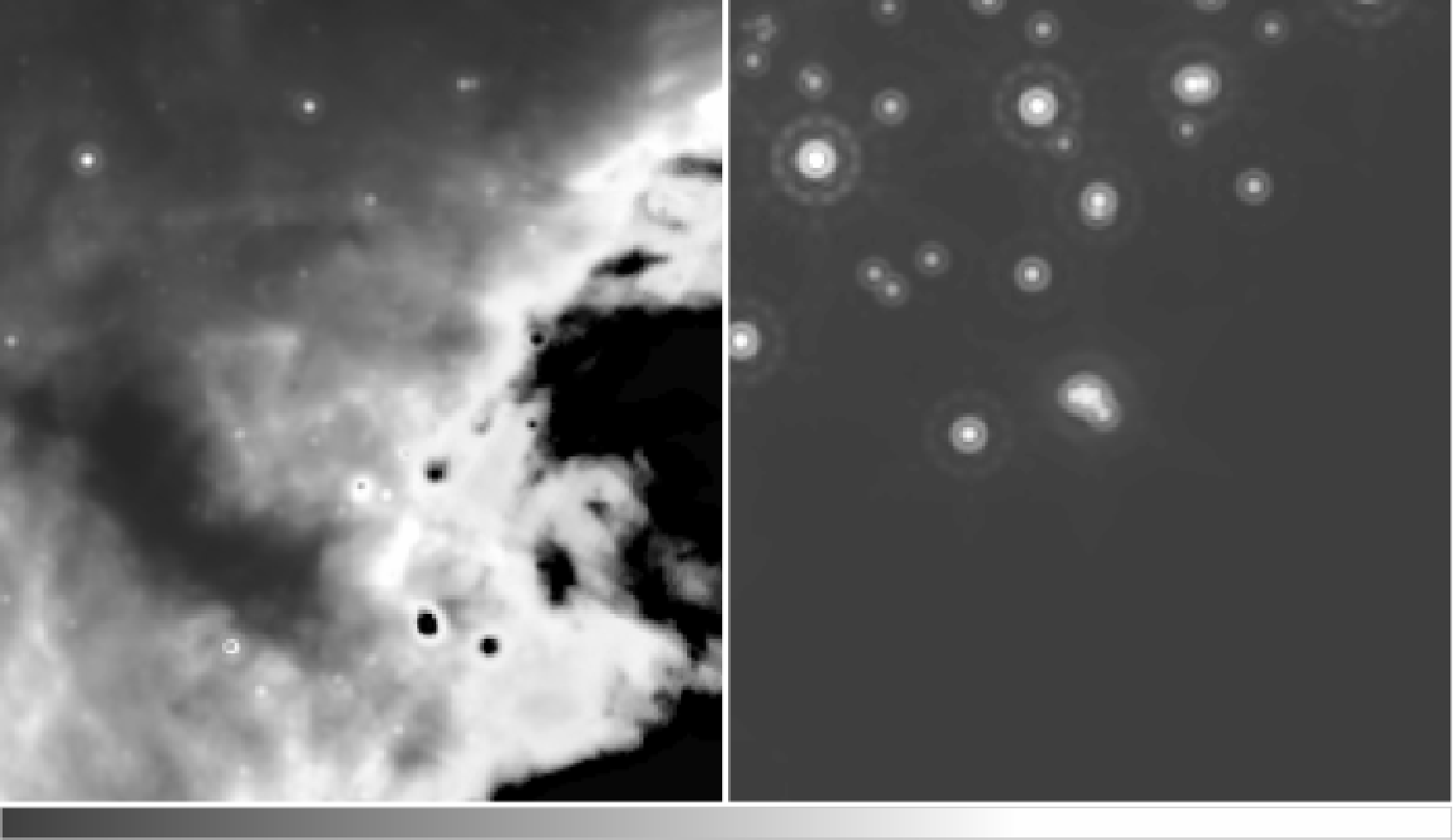}
\caption{On the left, the MIPS 24\,$\micron$ image of a $4\arcmin\times4\arcmin$
area near the GC at 17h45m59.6s, -28d43m25.7s.  On the right is the StarFinder
image of recovered sources.  In this case, the software mistakes
an area of bright diffuse emission as a series of point sources in the
center of the image.  Other point sources found in this area are much
brighter than their real counterparts (indicated by the size of the object
in the StarFinder image), implying that flux densities for point sources
in regions of bright diffuse emission may be unreliable.}
\end{figure}

\begin{figure}
\plotone{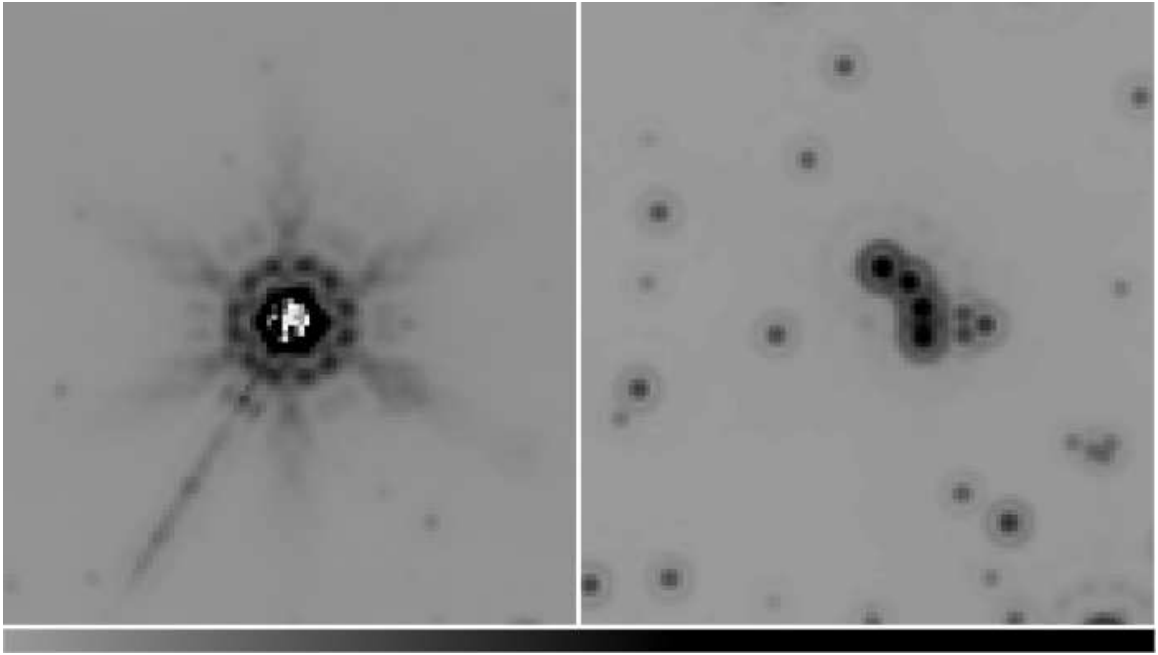}
\caption{On the left, the MIPS 24\,$\micron$ image of a saturated source
centered at 17h53m18.8s, -26d56m37.2s.  On the right is the StarFinder image 
of recovered sources.  The image on the right shows that the software mistakes 
the MIPS instrument image pattern
for a one-third ring of real objects.  The orientation of the image is along
the galactic plane, and the field of view is $4\farcm5$ on a side.}
\end{figure}

\begin{figure}
\plotone{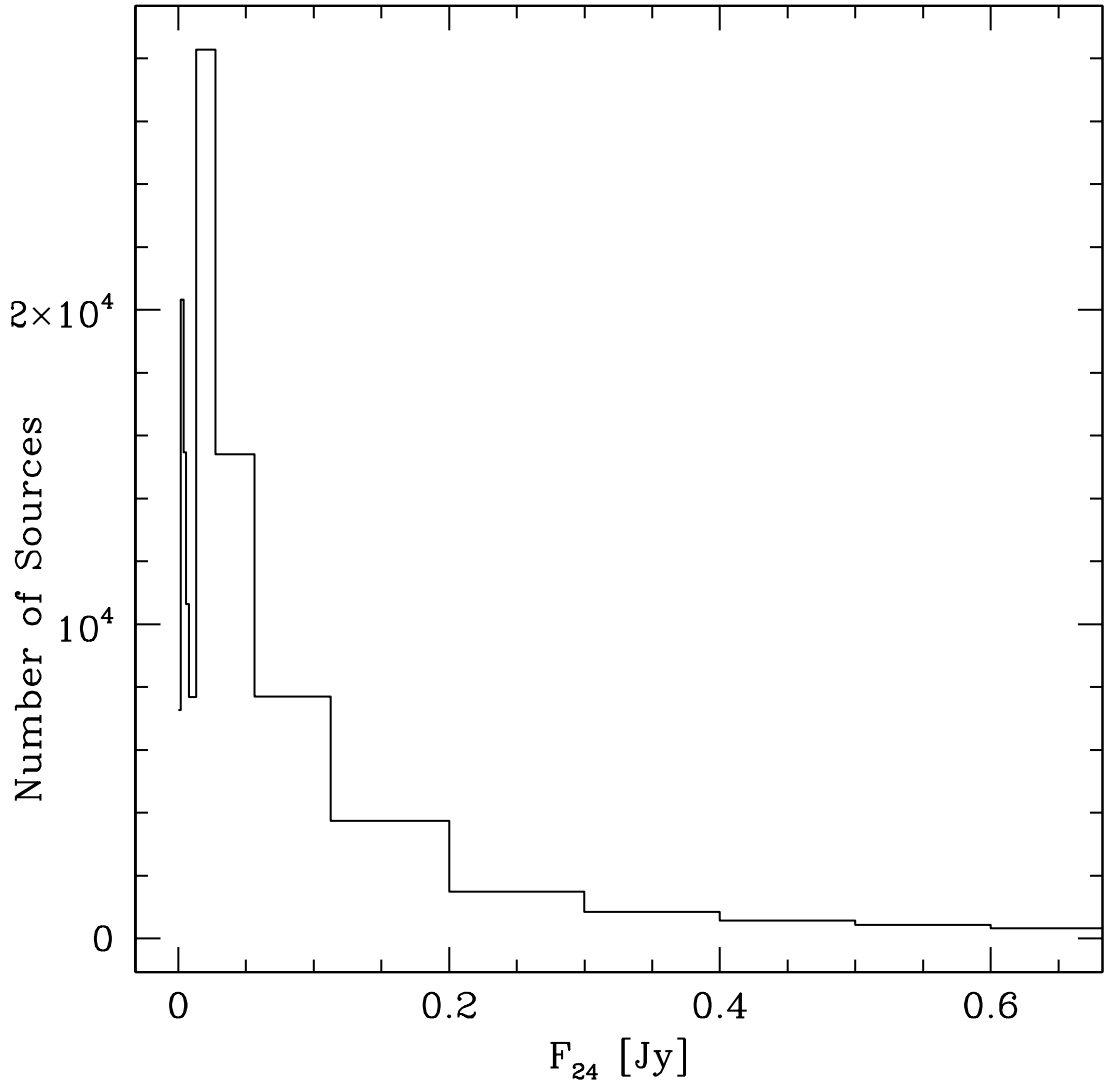}
\caption{Flux density distribution of sources in the 24\,$\micron$ image of the
GC.  The sources range in brightness from 0.6\,mJy to 9\,Jy.  The bin intervals
are as follows:  0-0.002\,Jy, 0.002-0.004\,Jy, 0.004-0.006\,Jy, 
0.006-0.008\,Jy, 0.008-0.01\,Jy, 0.01-0.0225, 0.0225-0.056\,Jy, 0.056-0.11\,Jy,
0.11-0.2\,Jy, 0.2-0.3\,Jy, 0.3-0.4\,Jy, 0.4-0.5\,Jy, 0.5-0.6\,Jy,$\ge0.6$\,Jy.}
\end{figure}

\begin{figure}
\plotone{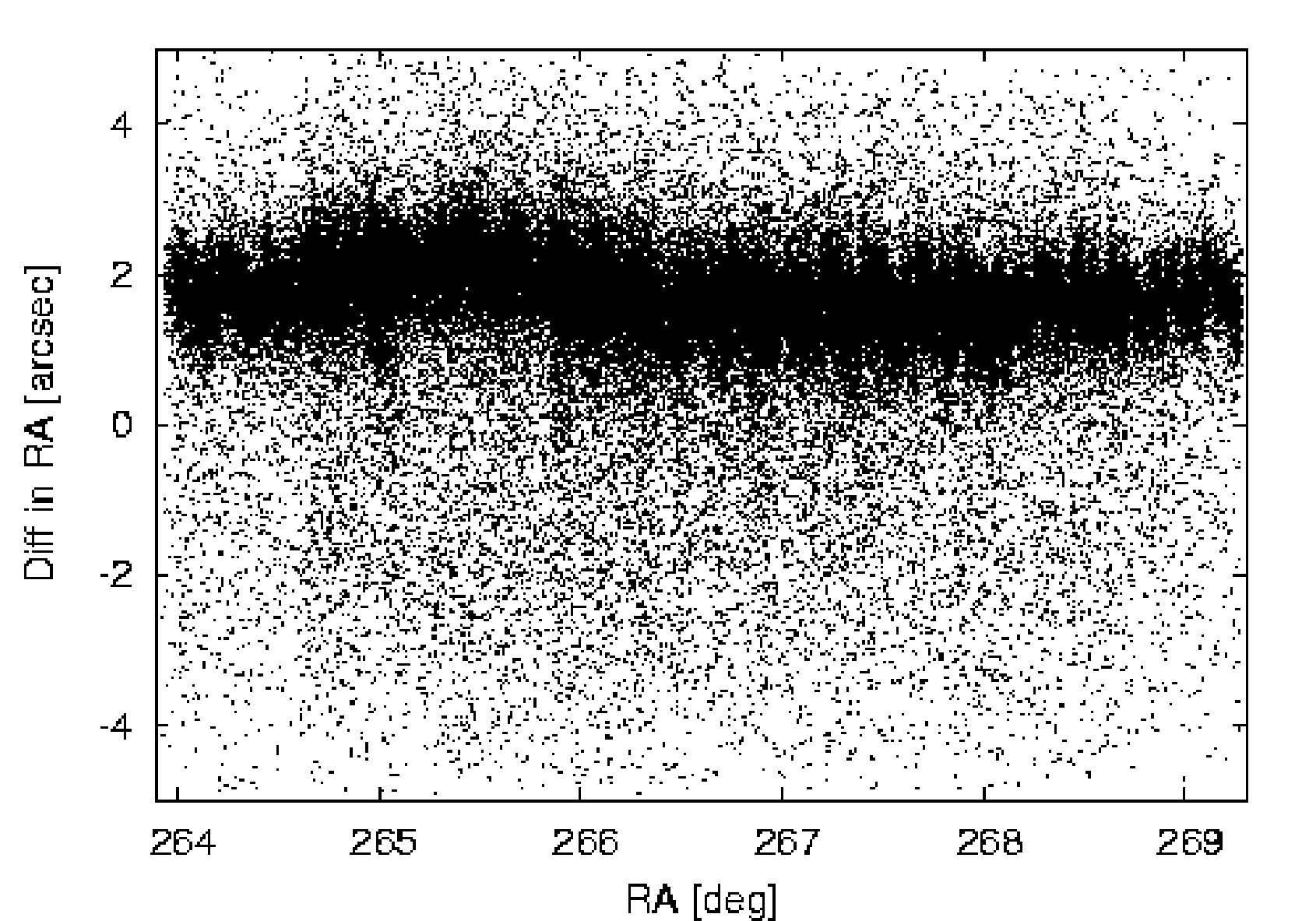}
\caption{The difference in coordinates between the
24\,$\micron$ and 2MASS sources as a function of the right ascension.  
Variations across the field were removed, with each of the 50 tiles of the
GC image having an individual correction which was applied to each source
within that tile.}
\end{figure}


\begin{references}

\reference{}Arendt, R.~G., et al.\ 2008, ArXiv e-prints, 804, arXiv:0804.4491 
\reference{}Armstrong, J.~T., \& Barrett, A.~H.\ 1985, \apjs, 57, 535 
\reference{}Bally, J., Stark, A.~A., Wilson, R.~W., \& Henkel, C.\ 1988, \apj, 324, 223 
\reference{}Blommaert, J.~A.~D.~L., van der Veen, W.~E.~C.~J., van Langevelde, H.~J., Habing, H.~J., \& Sjouwerman, L.~O. 1998, \aap, 329, 991
\reference{}Becklin, E.~E., \& Neugebauer, G.\ 1968, \apj, 151, 145 
\reference{}Campbell, M.~F., Niles, D.~W., Kanskar, M., Hoffmann, W.~F., \& Thronson, H.~A.\ 1985, Advances in Space Research, 5, 3
\reference{}Carey, S.~J., et al.\ 2005, Bulletin of the American Astronomical Society, 37, 1252 
\reference{}Carey, S.~J., et al.\ 2006, Bulletin of the American Astronomical Society, 38, 1023
\reference{}Cutri, R.~M., et al.\ 2003, The IRSA 2MASS All-Sky Point Source Catalog, NASA/IPAC Infrared Science Archive.~http://irsa.ipac.caltech.edu/applications/Gator/
\reference{}Diolaiti, E., Bendinelli, O., Bonaccini, D., Close, L.~M., Currie, D.~G., \& Parmeggiani, G.\ 2000, \procspie, 4007, 879 
\reference{}Engelbracht, C.~W. et al.\ 2007, accepted to \pasp
\reference{}Fazio, G.~G., et al.\ 2004, \apjs, 154, 10
\reference{}Gordon, K.~D., et al.\ 2005, \pasp, 117, 503
\reference{}Gordon, K.~D., et al.\ 2006, \apj, 638, L87
\reference{}Herrnstein, R.~M., \& Ho, P.~T.~P.\ 2005, \apj, 620, 287 
\reference{}Hines, D.~C. 2007, Bulletin of the American Astronomical Society, 211, 1210
\reference{}Jacoby, G.~H., \& Van de Steene, G. 2008, to be published in \aap
\reference{}Kessler, M. F., et al. 1996, \aap, 315, L27
\reference{}Lang, K.~R. 1992, "Astrophysical Data: Planets and Stars," Springer: New York
\reference{}Lis, D.~C., \& Carlstrom, J.~E.\ 1994, \apj, 424, 189 
\reference{}Meixner, M., et al.\ 2006, \aj, 132, 2268 
\reference{}Mill, J.~D., et al.\ 1994, Journal of Spacecraft and Rockets, 31, 900 
\reference{}Odenwald, S.~F., \& Fazio, G.~G.\ 1984, \apj, 283, 601
\reference{}Omont, A., et al.\ 2003, \aap, 403, 975 
\reference{}Papovich, C., et al.\ 2004, \apjs, 154, 70
\reference{}Pierce-Price, D., et al.\ 2000, \apjl, 545, L121 
\reference{}Price, S.~D., Egan, M.~P., Carey, S.~J., Mizuno, D.~R., \& Kuchar, T.~A.\ 2001, \aj, 121, 2819
\reference{}Ramirez, S.~V., Arendt, R.~G., Sellgren, K., Stolovy, S.~R., Cotera, A., Smith, H.~A., \& Yusef-Zadeh, F. 2008, \apjs, 175, 147
\reference{}Reach, W.~T. et al., 2006, \aj, 131, 1479
\reference{}Rieke, G.~H., \& Lebofsky, M.~J. 1985, \apj, 288, 618
\reference{}Rieke, G.~H., Telesco, C.~M., \& Harper, D.~A.\ 1978, \apj, 220, 556 
\reference{}Rieke, G.~H., et al.\ 2004, \apjs, 154, 25 
\reference{}Robitaille, T.~P., Cohen, M., Whitney, B.~A., Meade, M., Babler, B., Indebetouw, R., \& Churchwell, E.\ 2007, \aj, 134, 2099 
\reference{}Shenoy, S.~S., et al.\ 2007, American Astronomical Society Meeting Abstracts, 210, \#12.01 
\reference{}Skrutskie, M.~F., et al.\ 2006, \aj, 131, 1163
\reference{}Stolovy, S., et al.\ 2006, Journal of Physics Conference Series, 54, 176 
\reference{}Tokunaga, A.~T. 1999, in "Astrophysical Quantities," 4th Ed., Ed: Arthur Cox, Springer: New York
\reference{}Yusef-Zadeh, F., Hewitt, J.~W., \& Cotton, W.\ 2004, \apjs, 155, 421 
\reference{}Yusef-Zadeh, F., et al.\ 2008, American Astronomical Society Meeting Abstracts, 212, \#18.07 
\reference{}Zacharias, N., McCallon, H.~L., Kopan, E., \& Cutri, R.~M. 2003, 25th Meeting of the IAU, 16, 43

\end{references}
\end{document}